# Hydrogen–Diesel Dual-Fuel Engine Operation Using Real-Time GRU-Based Nonlinear Model Predictive Control

Alexander Winkler[a,*], Vasu Sharma[a], Julian Bedei[a], Edward Sperling[b], Charles Robert Koch[b], David Gordon[b], Jakob Andert[a]

[a] *Chair of Mechatronics in Mobile Propulsion, RWTH Aachen University, Forckenbeckstr. 4, Aachen, 52074, NRW, Germany*
[b] *Department of Mechanical Engineering, University of Alberta, 116 St & 85 Ave, Edmonton, T6G 2R3, AB, Canada*

## Abstract

Hydrogen-diesel dual-fuel (H2DF) combustion can reduce combustion-out $CO_2$ emissions but exhibits nonlinear cycle-to-cycle dynamics at high hydrogen energy shares (HES), while the $NO_x$–particulate matter (PM) trade-off must still be managed. This work evaluates nonlinear model predictive control (NMPC) with a gated recurrent-unit deep neural network dynamics model for transient H2DF control. Trained on 99,800 engine cycles, the model predicts indicated mean effective pressure, nitrogen oxides ($NO_x$), particulate matter (PM), and maximum pressure-rise rate. Experiments on a four-cylinder 4.46 L Cummins engine with hydrogen port injection in cylinder 1 only, naturally aspirated, at $1500\,\mathrm{min}^{-1}$, follow an unseen 4,900-engine-cycle trajectory. Compared with diesel-only operation under the production calibration, the NMPC reduces the load-tracking mean absolute error by 27.8% while reaching a mean HES of 39.7% (peak 55.1%); a high-hydrogen setting achieves 77.8% peak HES without measured constraint violations. The mean engine-out PM concentration is about 60% lower, while the mean engine-out $NO_x$ concentration increases by 105.1% without exhaust-gas recirculation. The controller tolerates feedback noise and model–plant mismatch from engine-speed variation, and it executes in 3 to 7 ms per engine cycle on low-cost embedded hardware within a modular rapid-prototyping toolchain. Learned-dynamics NMPC thus enables practical, real-time, constraint-aware transient H2DF control with substantial diesel substitution.

## 1. Introduction

Road transport remains a major contributor to global $CO_2$ emissions, and decarbonizing medium- and heavy-duty applications is particularly challenging due to high energy demand and long asset lifetimes [1, 2, 3]. While electrification is advancing, battery and infrastructure constraints still limit immediate rollout in many freight and off-road use cases [4, 5]. Hydrogen-based propulsion is therefore considered an important complementary pathway for near- and mid-term emission reduction [6, 7, 8, 9].

A practical transitional option is hydrogen-diesel dual-fuel (H2DF) combustion by retrofitting existing compression-ignition engines. By substituting a portion of diesel with port-injected hydrogen, H2DF can reduce soot and direct $CO_2$ emissions while preserving existing engine platforms and operational flexibility [10, 11, 12, 13, 14].

The literature on H2DF operation consistently reports strong particulate matter (PM) and $CO_2$ reductions with increasing hydrogen substitution, while $NO_x$ remains strongly load- and calibration-dependent [15, 16, 17, 18, 11, 19, 20]. Most published studies remain focused on steady-state operation and calibration-oriented control (maps, feedforward structures, or lightly tuned stock-engine-control-unit strategies), with limited evidence on closed-loop transient tracking under practical constraints [21, 22].

The present work builds directly on a progression of prior contributions by the authors' respective research groups. Norouzi et al. [23] introduced deep-neural-network (DNN)-based nonlinear model predictive control (NMPC), with a recurrent long short-term memory layer in the model, for diesel injection control on the same Cummins QSB 4.5L test bench. Further recurrent long short-term memory NMPC work, with full experimental validation on diesel emission control, is detailed in [24]. A rapid control prototyping workflow including low-cost hardware deployment was developed in [25]. This learned-model NMPC paradigm was subsequently extended to homogeneous-charge-compression-ignition low-temperature combustion [25] and, in [26], to thermal

*Corresponding author.
*Email addresses:* alexander.winkler@rwth-aachen.de (Alexander Winkler), vasu.sharma@rwth-aachen.de (Vasu Sharma), julian.bedei@rwth-aachen.de (Julian Bedei), ewsperli@ualberta.ca (Edward Sperling), bob.koch@ualberta.ca (Charles Robert Koch), dgordon@ualberta.ca (David Gordon), andert@mmp.rwth-aachen.de (Jakob Andert)

torque derating of electric machines, with each work introducing methodological extensions and experimental validation showing that the architecture transfers across different physical domains.

Mehnatkesh et al. [27] demonstrated real-time H2DF DNN-NMPC with a Bayesian-optimized feedforward model, four-variable injection control, explicit output constraints, and an experimental comparison against the production diesel controller. Further H2DF work of the authors' groups addresses dynamic emission analysis [28] and safe reinforcement-learning control [29]; the present study is complementary and concentrates on recurrent-model NMPC. More broadly, data-driven NMPC has shown promise for rapid control prototyping across various processes and systems, but experimental evidence for recurrent, cycle-to-cycle H2DF control and robustness to measurement noise and model–plant mismatch remains limited.

This work addresses this gap using a gated recurrent unit (GRU)-based deep-neural-network (DNN) dynamics model embedded in NMPC, hereafter referred to as GRU-NMPC, to investigate H2DF operation under practical constraints. The controller is validated on an unseen 4,900-engine-cycle transient trajectory and demonstrates load tracking at high hydrogen energy shares (HES) without measured constraint violations and real-time execution on low-cost embedded hardware. Relative to prior H2DF DNN-NMPC, the distinguishing contribution is the integration of recurrent cycle-to-cycle dynamics with embedded experiments that test robustness to measurement noise and model–plant mismatch.

Following this, the scientific contributions of this work are:

- Experimental demonstration of indicated mean effective pressure (IMEP) load tracking on an unseen 4,900-engine-cycle transient trajectory while sustaining high HES, with the resulting PM and $NO_x$ trade-off quantified against diesel-only operation.
- A constraint-aware GRU-NMPC formulation that integrates recurrent cycle-to-cycle dynamics, optimizes four actuator variables while bounding $NO_x$, PM and maximum pressure rise rate (MPRR), and is evaluated in embedded robustness experiments with measurement noise and engine-speed-induced model–plant mismatch.
- A modular rapid-prototyping workflow with real-time Raspberry Pi (RPi) execution in 3 to 7 ms per engine cycle, enabling controller updates without modifying the main engine-control software.

The paper structure follows the workflow summarized in Fig. 1. Section 2 describes the experimental setup, data generation, GRU-DNN dynamics-model identification, and NMPC formulation using `acados` [30]. Section 3 presents embedded experimental results for multiple controller variants and robustness investigations, Section 4 discusses them, and Section 5 summarizes the findings and outlook.

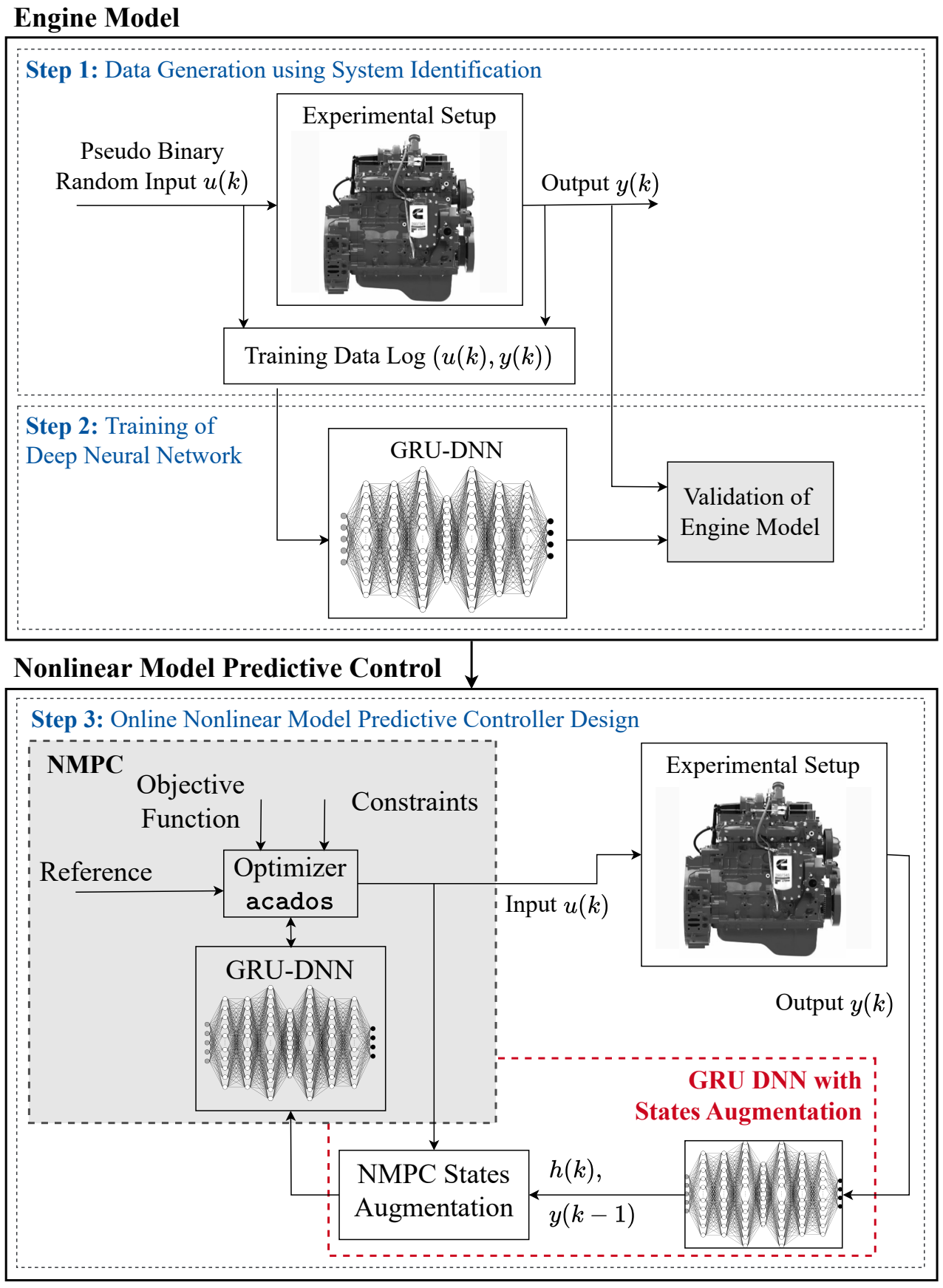

Figure 1: Workflow connecting H2DF operation, GRU-DNN dynamics modelling, NMPC design, and embedded deployment.

## 2. Methodology

### *2.1. Experimental Setup*

A modified four-cylinder Cummins QSB 4.5L diesel engine with a compression ratio of 17.2:1 serves both for data collection to train the GRU-DNN dynamics model and for experimental validation of the GRU-NMPC. Further details on H2DF combustion characterization for this platform are available in [32].

A schematic of the experimental setup is shown in Fig. 2. Table 1 describes the engine parameters in detail. This engine is a direct-injection common-rail diesel engine, now naturally aspirated after removal of its original turbocharger. The main hardware modification to the engine is the addition of a hydrogen port fuel injection (HANA H2200A) in the intake manifold of cylinder 1. The exhaust gas leaving cylinder 1 has been separated from the other three cylinders to allow for independent analysis. Only cylinder 1 is fuelled with hydrogen; all $p_{\mathrm{mi}}$, $dp_{\mathrm{max}}$, $c_{\mathrm{NOx}}$ and $c_{\mathrm{PM}}$ values in this work refer to cylinder 1. Hydrogen is stored in K-tanks with a nominal pressure of 200 bar. A single-stage high flow hydrogen rated stainless steel pressure regulator set to supply hydrogen at a static pressure of 8.3 bar is utilized. Additionally, the production engine control unit has been replaced to allow for fully flexible engine control development in `MATLAB/Simulink`. The rapid-control-prototyping system uses a dSPACE MicroAutoBox II (MABX) and RapidPro power stage. The NMPC optimization runs on the RPi 400 with an ARM Cortex-A72 processor and communicates with the main engine controller on the MABX via the User Datagram Protocol over Ethernet; the hardware and communication path are shown in Fig. 2. The hydrogen injector is originally designed as a compressed-natural-gas port injector for industrial engines with a cylinder volume of approximately 2.5L. It is noteworthy that this engine is not equipped with an exhaust gas recirculation (EGR) system, which is known to reduce $NO_x$ emissions by lowering peak combustion temperatures [33].

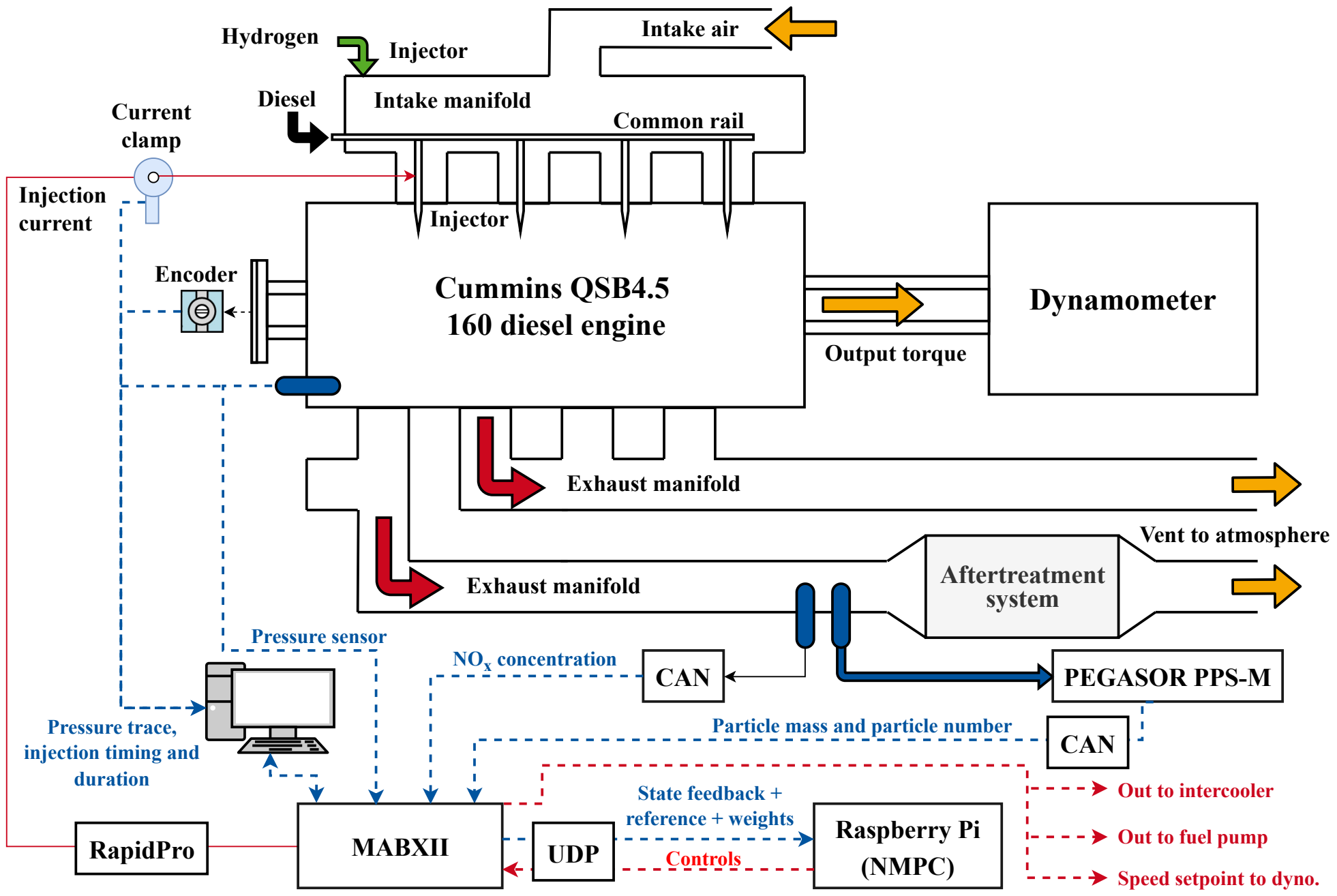

Figure 2: Experimental setup of the modified Cummins QSB 4.5L diesel engine, including the Raspberry Pi (RPi) optimization target and its communication path to the engine controller. The engine is retrofitted with hydrogen port fuel injection at cylinder 1; the underlying test-bench configuration is described in [31].

In-cylinder pressure is measured using a Kistler 6124A piezoelectric pressure transducer, which is mounted between the exhaust valves. Using the measured pressure of cylinder 1, $p_{mi}$ and $dp_{max}$ can be calculated on a Xilinx Kintex-7 field-programmable gate array. Additional information can be found in [34].

Engine-out particle-mass concentration $c_{PM}$ and $NO_x$ concentration $c_{NO_x}$ are measured directly after the exhaust valve of cylinder 1, without any additional aftertreatment. A Pegasor PPS-M particle sensor is used to measure engine-out particle-mass concentration $c_{PM}$ in mg/m$^3$. The PPS-M has a transient response time of 0.2 s. Its performance is sensitive to particle-size distribution and maintenance quality, with measurement errors of up to 50% reported in the literature [35]. The Pegasor data is transmitted via a controller-area network and logged by the control host computer. The $NO_x$ volume concentration in parts per million (ppm) in the exhaust stream is measured using an ECM NOxCANt measurement module with a calibrated $NO_x$ sensor. Combustion phasing and center of heat release are outside the scope of this work and were not considered during engine operation and analysis. Crank-angle positions are reported in crank angle degrees (CAD) relative to top dead centre (TDC); bTDC and aTDC denote before and after TDC, respectively. In summary, there are four engine-out (cylinder 1) outputs measured and relevant for this work as depicted in Table 2.

Table 1: Cummins QSB 4.5 Tier 3 engine parameters, as tested.

| **Parameter** | **Value** |
|---|---|
| Total displacement | 4.46 L |
| Number of cylinders | 4 |
| Bore | 107 mm |
| Stroke | 124 mm |
| Connecting rod length | 192 mm |
| Compression ratio | 17.2:1 |
| Piston protrusion | 0.43 mm |
| Head gasket thickness | 1.6 mm |
| Valve train | Pushrod |
| No. of valves (In/Ex) | 2/2 |
| No. of camshafts | 1 |
| Max. valve lift (In/Ex) | 8 mm/8 mm |
| Valve angle (In/Ex) | 0°/0° |
| Valve diameter (In/Ex) | 33 mm/33 mm |
| Combustion chamber type | Bowl in piston |
| Injection type | Direct |
| Injection pressure | 300 to 1500 bar |
| Spray angle | 124° |
| Injector holes | 8 holes |
| Injection plane | Single |
| Injector actuation | Solenoid |

The following operating parameters are externally controlled and held constant: engine coolant temperature, oil temperature, engine speed, and hydrogen fuel pressure. To reduce the control-problem complexity, diesel injection is limited to pre and main events, and the following quantities are fixed: pre-injection duration $t_{\mathrm{Pre}} = 0.17\,\mathrm{ms}$, corresponding to 3 mg of diesel fuel; diesel rail pressure $p_{\mathrm{Rail}} = 970\,\mathrm{bar}$; hydrogen start of injection $\alpha_{\mathrm{H2}} = 360\,\mathrm{CAD\,bTDC}$; and naturally aspirated intake pressure. At $1500\,\mathrm{min}^{-1}$, one four-stroke engine cycle corresponds to 80 ms. As shown in Fig. 3, the combustion metrics are available by 165 CAD aTDC and the next injection setpoint is required at 330 CAD aTDC. The available online calculation time is therefore $t_{\mathrm{calc}} \approx 18\,\mathrm{ms}$, corresponding to 165 CAD. Subsequently, there are four engine controls available to the NMPC, namely the main-diesel duration of injection (DOI), the pre-to-main (P2M) interval, the main-diesel start of injection (SOI), and the hydrogen DOI, as stated in Table 3.

The hydrogen energy share (HES) reported throughout is calculated per engine cycle from the injected fuel masses,

$$\mathrm{HES}(k) = \frac{m_{\mathrm{H2}}(k)\, h_{\mathrm{H2}}}{m_{\mathrm{H2}}(k)\, h_{\mathrm{H2}} + m_{\mathrm{D}}(k)\, h_{\mathrm{D}}}, \tag{1}$$

where $m_{\mathrm{H2}}$ and $m_{\mathrm{D}}$ are the hydrogen and total diesel (pre plus main injection) masses per cycle, and $h_{\mathrm{H2}} = 120\,\mathrm{MJ/kg}$ and $h_{\mathrm{D}} = 42.5\,\mathrm{MJ/kg}$ are the lower heating values. The masses are not measured; they are calculated from the commanded injection durations with injector flow maps for hydrogen (regulated pressure 8.3 bar) and diesel (rail pressure 970 bar). HES is therefore a calculated quantity whose uncertainty follows that of the maps, which is not quantified here. The diesel energy share is $\mathrm{DES} = 1 - \mathrm{HES}$ [32].

The diesel pre-injection timing $\alpha_{\mathrm{Pre}}$ depends on the P2M time $t_{\mathrm{P2M}}$. Furthermore, $t_{\mathrm{P2M}}$ is calculated using the engine speed $n_{\mathrm{Eng}}$ in $\mathrm{min}^{-1}$ as:

$$t_{\mathrm{P2M}}(k) = \left( \frac{\alpha_{\mathrm{Pre}}(k) - \alpha_{\mathrm{Main}}(k)}{6 \cdot n_{\mathrm{Eng}}(k)} - t_{\mathrm{Pre}}(k) \right). \tag{2}$$

Using $t_{\mathrm{P2M}}$ instead of $\alpha_{\mathrm{Pre}}$ simplifies the control task by defining pre-injection timing relative to the main injection [32].

Table 2: Measured engine output variables for cylinder 1.

| **Output variable** | **Symbol** | **Unit** |
|---|---|---|
| Indicated mean effective pressure | $p_{\mathrm{mi}}$ | bar |
| $\mathrm{NO_x}$ | $c_{\mathrm{NOx}}$ | ppm |
| Particulate matter | $c_{\mathrm{PM}}$ | $\mathrm{mg/m^3}$ |
| Maximum pressure rise rate | $dp_{\max}$ | bar/CAD |

Table 3: Engine control inputs for cylinder 1.

| **Input variable** | **Symbol** | **Unit** |
|---|---|---|
| Main-diesel injection duration | $t_{\mathrm{Main}}$ | ms |
| Pre-to-main interval | $t_{\mathrm{P2M}}$ | µs |
| Main-diesel start of injection | $\alpha_{\mathrm{Main}}$ | CAD bTDC |
| Hydrogen injection duration | $t_{\mathrm{H2}}$ | ms |

### *2.2. Process Characteristics and Control-Relevant Variables*

H2DF combustion differs markedly from conventional diesel operation. The onset of combustion is slightly delayed but occurs close to TDC, after which hydrogen's high flame speed produces a rapid heat release. This results in significantly higher $dp_{\max}$ and peak cylinder pressures, increasing mechanical stress and sensitivity to operating conditions. Compared to diesel's longer and slower burn duration, H2DF combustion is more concentrated around TDC, resembling a closer-to-constant-volume process. Consequently, $p_{\mathrm{mi}}$ remains similar despite sharper pressure traces, while the indicated efficiency increases because less fuel energy is required to achieve the same $p_{\mathrm{mi}}$. Rising HES levels amplify these effects: low HES levels produce moderate pressure increases, whereas higher HES levels lead to steep pressure rises and elevated peak cylinder pressures, reflecting faster combustion and stronger thermal loading. H2DF operation also exhibits increased cycle-to-cycle variability due to hydrogen's ignition sensitivity and mixture dynamics, contributing to combustion instability and noise. These characteristics highlight the need for precise, fast control systems that account for cycle-to-cycle effects while managing nonlinear dual-fuel behavior, performance, pressure, durability, and emission constraints.

The injection timings and control setpoints over one combustion cycle are visualized in Fig. 3. Here and in the following controller equations, $k$ denotes the engine-cycle index. The controls for the NMPC are represented in compact form as

$$u(k) = \Big[ t_{\mathrm{Main}}(k) \quad t_{\mathrm{P2M}}(k) \quad \alpha_{\mathrm{Main}}(k) \quad t_{\mathrm{H2}}(k) \Big]^{\mathsf{T}}. \tag{3}$$

While $u(k)$ represents the physical actuator commands sent to the engine, the NMPC optimizes over the control changes $\Delta u(k) = u(k) - u(k-1)$, namely the input increments rather than their absolute values. This formulation allows direct penalization of actuator change rates in the cost function and simplifies rate-of-change constraint specification.

The corresponding output vector is written as

$$y(k) = \Big[ p_{\mathrm{mi}}(k) \quad c_{\mathrm{NOx}}(k) \quad c_{\mathrm{PM}}(k) \quad dp_{\max}(k) \Big]^{\mathsf{T}}. \tag{4}$$

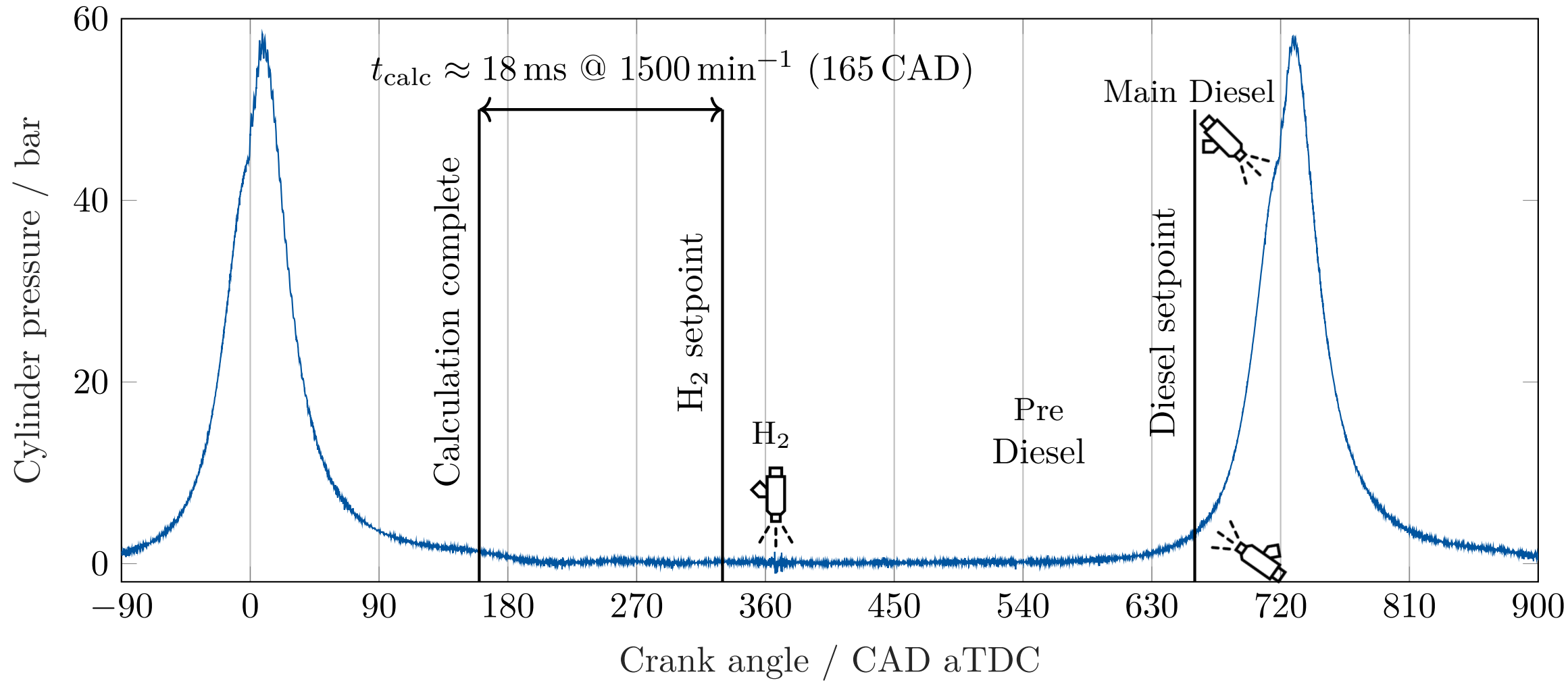

Figure 3: Cylinder pressure trace with injection timing events at engine speed $n_{Eng} = 1500\,\text{min}^{-1}$. Crank-angle positions use crank angle degrees (CAD) relative to top dead centre (TDC); after TDC is abbreviated aTDC.

### *2.3. GRU-DNN Dynamics Model*

The controller uses a data-driven GRU-DNN dynamics model to capture the strongly nonlinear and engine-cycle-dependent H2DF combustion behavior. Its integration as the dynamics model of the GRU-NMPC is visualized in Fig. 1. The bundled model input vector $X_{DNN}(k)$ collects the controls of Eq. 3 together with the load feedback $p_{mi}(k-1)$, i.e. $X_{DNN}(k) = [u(k)^\top \; p_{mi}(k-1)]^\top \in \mathbb{R}^5$, while the bundled model output vector $Y_{DNN}(k)$ comprises the outputs defined in Eq. 4. The GRU-DNN architecture is shown in Fig. 4; it combines fully connected (FC) layers with a GRU core to represent dynamic effects while keeping the state dimension suitable for real-time optimization.

Model training is performed on measured engine data generated by pseudo-random binary-sequence excitation of the four control inputs, collected in three individual measurement runs (about 130 min in total). Each measurement run is split chronologically into training, validation, and test subsets using an 80%/15%/5% split, without shuffling, to preserve the temporal structure required by the GRU; the corresponding subsets from all runs are then concatenated for model development. The selected network is the best-validation model, and the final training setup is summarized in Table A.1. Prediction performance on the held-out test set (the final 5% of each run) is summarized in Table 4 and in the corresponding measured-versus-predicted output trajectories in Fig. A.3. Prediction errors are reported as root mean square error (RMSE) and normalized root mean square error (NRMSE); the clipped variant $\text{NRMSE}_{clip}$ is used for the model metrics. The GRU-DNN dynamics model provides high accuracy for load and $NO_x$ prediction ($p_{mi}$: RMSE 0.23 bar, $\text{NRMSE}_{clip}$ 3.0%; $c_{NOx}$: RMSE 40 ppm, $\text{NRMSE}_{clip}$ 4.0%), with moderate accuracy for $dp_{max}$ ($\text{NRMSE}_{clip}$ 12.1%) and lower but acceptable accuracy for $c_{PM}$ ($\text{NRMSE}_{clip}$ 9.2%). The comparatively lower $c_{PM}$ prediction accuracy should be interpreted in light of the PPS-M measurement uncertainty discussed in Section 2.1, which limits the fidelity of the corresponding target data. Here $\text{NRMSE}_{clip}$ is normalized by the output range after values are clipped to the 1st and 99th percentiles (two-sided winsorization), limiting the influence of sensor outliers on the normalization denominator [36]. The test segments come from the same three runs and the same excitation and are therefore not an independent validation of the model; the closed-loop 4,900-engine-cycle trajectory of Section 3, which was not used for training, serves as the independent check.

### *2.4. GRU-NMPC Formulation*

NMPC determines control actions by repeatedly solving a finite-horizon optimal control problem (OCP) online. At each sampling instant the OCP is initialised with the current augmented state (measured outputs, last controls and propagated GRU state, see below), only the first element of the optimal input sequence is applied, and the horizon is shifted forward. This receding-horizon principle provides feedback while handling multivariable objectives and constraints explicitly within the optimisation. [37] It is therefore well suited to transient H2DF operation, where competing objectives are coupled through the same actuators: the objective formulated here is load tracking subject to simultaneous moderation of engine-out emissions and the maximum pressure-rise rate. Solving this OCP requires a dynamics model that maps the applied inputs to the constrained outputs over the horizon. Fig. 1 shows the underlying principle: the GRU-DNN dynamics model takes the place of a physics-based model and supplies the state predictions propagated across the horizon at each sampling instant. Its dynamics therefore have to be formulated in a form suitable for embedding into the optimisation.

Table 4: GRU-DNN prediction metrics on the held-out test data. $\text{NRMSE}_{clip}$ is normalized by the output range after two-sided winsorization at the 1st and 99th percentiles.

| **Output variable** | **RMSE** | $\text{NRMSE}_{clip}$ |
|---|---|---|
| $p_{mi}$ | 0.23 bar | 3.0% |
| $c_{NOx}$ | 40 ppm | 4.0% |
| $c_{PM}$ | 0.14 mg/m$^3$ | 9.2% |
| $dp_{max}$ | 1.56 bar/CAD | 12.1% |

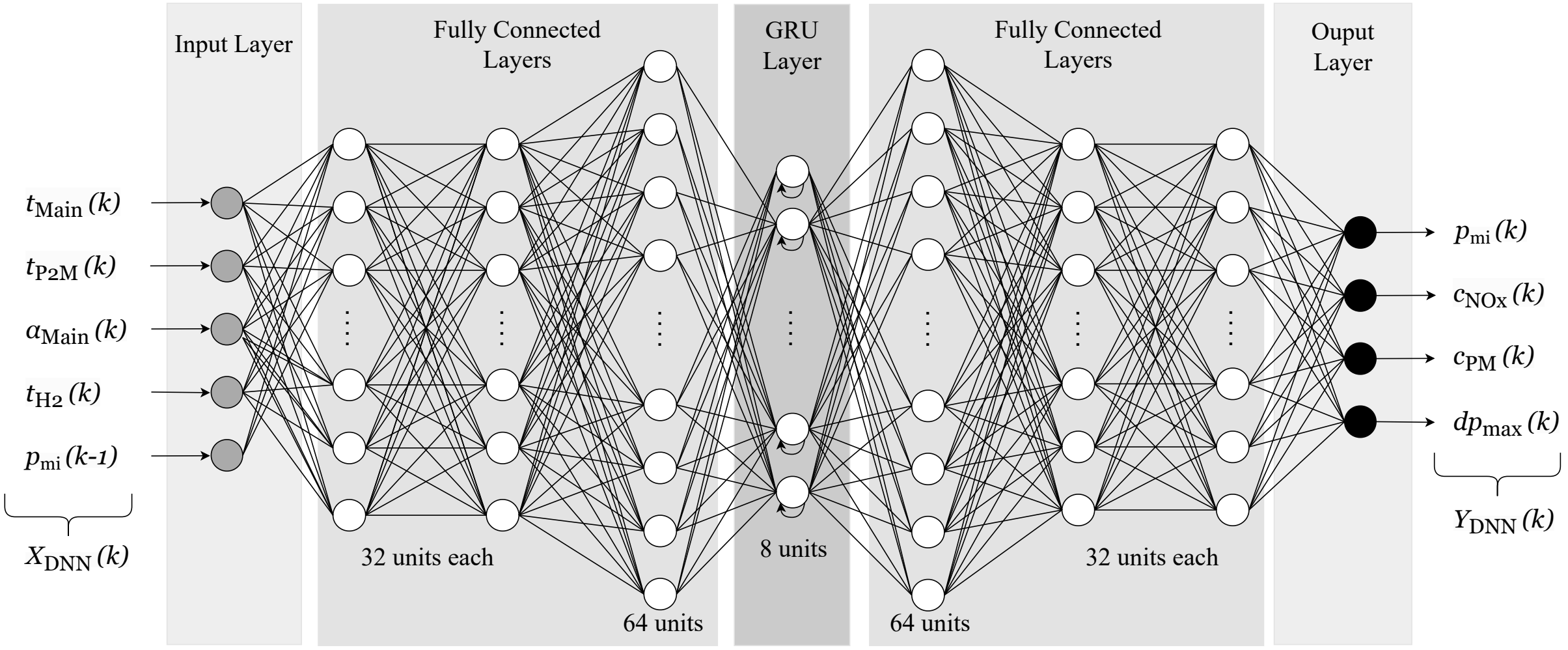

Figure 4: GRU-DNN dynamics-model architecture with load feedback $p_{\mathrm{mi}}(k-1)$ and 7,900 total learnable parameters. Fully connected layers with rectified linear unit (ReLU) activations surround the GRU layer. $X_{\mathrm{DNN}}(k)$ and $Y_{\mathrm{DNN}}(k)$ denote the bundled model input and output vectors, and $h(k)$ is the recurrent hidden state.

A detailed computational graph of the deployed network is provided in Fig. A.1. The DNN comprises a GRU layer enclosed by FC encoder and decoder layers with ReLU activation functions. The forward propagation through the three input FC layers is

$$z_{\mathrm{FC1}}(k) = \mathrm{ReLU}\big(W_{\mathrm{FC1}}^{\top} X_{\mathrm{DNN}}(k) + b_{\mathrm{FC1}}\big)\,, \tag{5a}$$
$$z_{\mathrm{FC2}}(k) = \mathrm{ReLU}\big(W_{\mathrm{FC2}}^{\top} z_{\mathrm{FC1}}(k) + b_{\mathrm{FC2}}\big)\,, \tag{5b}$$
$$z_{\mathrm{FC3}}(k) = \mathrm{ReLU}\big(W_{\mathrm{FC3}}^{\top} z_{\mathrm{FC2}}(k) + b_{\mathrm{FC3}}\big)\,. \tag{5c}$$

The recurrent update is

$$z(k) = \sigma\big(W_{u,z}^{\top} z_{\mathrm{FC3}}(k) + W_{h,z}^{\top} h(k-1) + b_z\big)\,, \tag{6a}$$
$$r(k) = \sigma\big(W_{u,r}^{\top} z_{\mathrm{FC3}}(k) + W_{h,r}^{\top} h(k-1) + b_r\big)\,, \tag{6b}$$
$$\hat{h}(k) = \tanh\big(W_{u,h}^{\top} z_{\mathrm{FC3}}(k) + W_{h,h}^{\top}(r(k) \odot h(k-1)) + b_h\big)\,, \tag{6c}$$
$$h(k) = (1 - z(k)) \odot h(k-1) + z(k) \odot \hat{h}(k). \tag{6d}$$

The three output FC layers then decode the hidden state:

$$z_{\mathrm{FC4}}(k) = \mathrm{ReLU}\big(W_{\mathrm{FC4}}^{\top} h(k) + b_{\mathrm{FC4}}\big)\,, \tag{7a}$$
$$z_{\mathrm{FC5}}(k) = \mathrm{ReLU}\big(W_{\mathrm{FC5}}^{\top} z_{\mathrm{FC4}}(k) + b_{\mathrm{FC5}}\big)\,, \tag{7b}$$
$$y(k) = W_{\mathrm{FC6}}^{\top} z_{\mathrm{FC5}}(k) + b_{\mathrm{FC6}}. \tag{7c}$$

Here, $W_{\mathrm{FC}i}$ and $b_{\mathrm{FC}i}$ are the weight matrix and bias vector of FC layer $i \in \{1, \ldots, 6\}$. The matrices $W_{u,(z,r,h)}$ map the encoded input $z_{\mathrm{FC3}}(k)$, while $W_{h,(z,r,h)}$ map the previous hidden state $h(k-1)$. The variables $z(k)$, $r(k)$, and $\hat{h}(k)$ denote the update gate, reset gate, and candidate state, respectively, and $\odot$ denotes element-wise multiplication.

Grouping the model into input FC, GRU, and output FC blocks gives

$$z_{\mathrm{FC3}}(k) = f_{\mathrm{FC,in}}(X_{\mathrm{DNN}}(k))\,, \tag{8a}$$
$$h(k) = f_{\mathrm{GRU}}(h(k-1), z_{\mathrm{FC3}}(k))\,, \tag{8b}$$
$$y(k) = f_{\mathrm{FC,out}}(h(k))\,. \tag{8c}$$

Here, $f_{\mathrm{FC,in}}$, $f_{\mathrm{GRU}}$, and $f_{\mathrm{FC,out}}$ summarize Eqs. 5, 6, and 7, respectively.

Eliminating the intermediate vector $z_{\mathrm{FC3}}(k)$ and defining the model state $x(k) = h(k-1)$ yields the nonlinear state-space model used by the NMPC:

$$x(k+1) = f_{\mathrm{FC,in:GRU}}(x(k), u(k)) = f(x(k), u(k))\,, \tag{9a}$$
$$y(k) = f_{\mathrm{FC,out}}(x(k+1)) = g(x(k), u(k))\,. \tag{9b}$$

In this compact representation, $f$ and $g$ denote the nonlinear state-transition and output-mapping functions, respectively. Here, $f_{\mathrm{FC,in:GRU}}$ combines the input FC layers and the GRU layer, while $f_{\mathrm{FC,out}}$ represents the output FC layers. The engine-cycle indexing is selected such that $y(k)$ reflects the influence of the current control input $u(k)$ through the updated recurrent state $x(k+1)$. Thus, the equations expose the dependence on both the previous hidden state and the current actuator input [38]. Eqs. 9 and 10 write the model input as $u(k)$ for brevity. In the deployed model the load feedback $p_{\mathrm{mi}}(k-1)$ is a fifth input. It is the load component of the previous model output, which the augmented NMPC state carries together with the eight GRU hidden states and the last controls, $\tilde{x} \in \mathbb{R}^{16}$. At the first stage of the horizon $p_{\mathrm{mi}}(k-1)$ is the measured value; at later stages it is the model's own prediction.

To improve control stability and reduce oscillations in the outputs, the gradients of the controls $\Delta u(k) = u(k) - u(k-1)$ are introduced as additional inputs [39]. This modification enables the penalization of both absolute inputs and their rate of change within the cost function. The weights therefore trade tracking and emission performance against actuator-magnitude and actuator-rate penalties.

Following this, using Eq. 9, the NMPC state vector $\tilde{x}(k)$ and NMPC output vector $\tilde{y}(k)$ are thus defined as:

$$\underbrace{\begin{bmatrix} x(k+1) \\ u(k) \end{bmatrix}}_{\tilde{x}(k+1)} = \underbrace{\begin{bmatrix} f\big(x(k),\, u(k-1) + \Delta u(k)\big) \\ u(k-1) + \Delta u(k) \end{bmatrix}}_{\tilde{f}\left(\tilde{x}(k),\, \Delta u(k)\right)}, \tag{10a}$$

$$\underbrace{\begin{bmatrix} y(k) \\ u(k-1) \end{bmatrix}}_{\tilde{y}(k)} = \underbrace{\begin{bmatrix} g\big(x(k),\, u(k-1)\big) \\ u(k-1) \end{bmatrix}}_{\tilde{g}(\tilde{x}(k))}, \tag{10b}$$

where $\tilde{f}$ and $\tilde{g}$ are the NMPC's augmented nonlinear state transition function and output mapping function, respectively.

At each engine cycle $k$, the NMPC solves a finite-horizon, discrete-time OCP over a prediction horizon of $N$ engine cycles. The optimisation is written directly in the H2DF-specific OCP-structured nonlinear-programming form using the augmented dynamics from Eq. 10. The index sets are defined as $\mathcal{I}_N = \{0, \ldots, N\}$ and $\mathcal{I}_{N-1} = \{0, \ldots, N-1\}$. Here, $i$ denotes the prediction-horizon index. The decision variables are the manipulated input increments $\tilde{u}_i = \Delta u_i$, augmented states $\tilde{x}_i$, predicted outputs $\tilde{y}_i$, and non-negative slack variables $s_i = (s_{\min,\tilde{x},i},\, s_{\max,\tilde{x},i})$. Soft constraints on the output bounds are introduced to ensure OCP feasibility, yielding:

$$\begin{aligned}
\min_{\substack{\tilde{u}_0,\ldots,\tilde{u}_{N-1} \\ \tilde{x}_0,\ldots,\tilde{x}_N \\ \tilde{y}_0,\ldots,\tilde{y}_N \\ s_0,\ldots,s_N}} \quad & \sum_{i=0}^{N} J_{\tilde{x}}\big(\tilde{x}_i, s_i\big) + \sum_{i=0}^{N-1} J_{\tilde{u}}\big(\tilde{u}_i\big) \\
\text{s.t.} \quad & \tilde{x}_0 = \tilde{x}(k), \\
& \tilde{x}_{i+1} = \tilde{f}\big(\tilde{x}_i, \tilde{u}_i\big) \quad \forall i \in \mathcal{I}_{N-1}, \\
& \tilde{y}_i = \tilde{g}\big(\tilde{x}_i\big) \quad \forall i \in \mathcal{I}_N, \\
& \tilde{u}_{\min} \le F_{\tilde{u}}\, \tilde{u}_i \quad \forall i \in \mathcal{I}_{N-1}, \\
& F_{\tilde{u}}\, \tilde{u}_i \le \tilde{u}_{\max} \quad \forall i \in \mathcal{I}_{N-1}, \\
& F_{\tilde{x}}\, \tilde{x}_i - F_{s,\tilde{x}}\, s_{\max,\tilde{x},i} \le \tilde{x}_{\max} \quad \forall i \in \mathcal{I}_N, \\
& F_{\tilde{x}}\, \tilde{x}_i + F_{s,\tilde{x}}\, s_{\min,\tilde{x},i} \ge \tilde{x}_{\min} \quad \forall i \in \mathcal{I}_N, \\
& \Big[s_{\min,\tilde{x},i},\, s_{\max,\tilde{x},i}\Big]^{\mathsf{T}} \ge 0 \quad \forall i \in \mathcal{I}_N.
\end{aligned} \tag{11}$$

Here, $\tilde{f}$ is the augmented state transition, $\tilde{g}$ is the output mapping, and $J_{\tilde{x}}(\cdot)$ and $J_{\tilde{u}}(\cdot)$ collect the state, slack, and input-rate cost terms detailed below. The prediction horizon length $N$ defines how many engine cycles the NMPC predicts into the future; the implemented controller uses $N = 3$. $F_{\tilde{u}}$ and $F_{\tilde{x}}$ select and scale constrained input and state components, while $F_{s,\tilde{x}}$ maps the slack components to the corresponding state bounds in Eq. 11. Only the first optimized input increment, $\tilde{u}_0$, is applied to the engine.

At every engine cycle the OCP is initialised with the measured outputs, the last applied controls and the GRU state propagated by the model from the previous cycle. As the four measured outputs re-enter the optimisation at every cycle, the predictions are re-anchored to the engine without a separate state estimator; the tolerance of the closed loop to model–plant mismatch is examined in Section 3.1.

Following the control objective, the state and slack cost $J_{\tilde{x}}\big(\tilde{x}_i, s_i\big)$ and the input-rate cost $J_{\tilde{u}}\big(\tilde{u}_i\big)$ are defined using a linear least squares formulation. The resulting finite-horizon cost is

$$\begin{aligned}
J = \sum_{i=0}^{N} \Big[ & \underbrace{\|p_{\mathrm{mi},i} - p_{\mathrm{mi,ref},i}\|^2_{q_{p,\mathrm{mi}}}}_{\text{Load tracking}} + \underbrace{\|dp_{\max,i}\|^2_{q_{dp,\max}}}_{\text{Pressure-rise moderation}} \\
& + \underbrace{\|c_{\mathrm{NOx},i}\|^2_{q_{c,\mathrm{NOx}}} + \|c_{\mathrm{PM},i}\|^2_{q_{c,\mathrm{PM}}}}_{\text{Emissions minimization}} \\
& + \underbrace{\|t_{\mathrm{Main},i}\|^2_{q_{t,\mathrm{Main}}} + \|t_{\mathrm{H2},i}\|^2_{q_{t,\mathrm{H2}}}}_{\text{Fuel consumption and } H_2 \text{ share}} \\
& + \underbrace{\|s_i\|^2_{q_{s,2}} + \|s_i\|^1_{q_{s,1}}}_{\text{Constraint violation}} \Big] + \sum_{i=0}^{N-1} \underbrace{\|\Delta u_i\|^2_{r_{\tilde{u}}}}_{\text{Input-rate moderation}} .
\end{aligned} \tag{12}$$

The weights $q$ and $r_{\tilde{u}}$ are the diagonal entries of positive semi-definite weighting matrices for the output/state and input-rate terms, respectively; the GRU hidden states carry zero weight. All physical and GRU-DNN inputs, outputs, references, and bounds in the OCP are normalized using fixed physical ranges. All constraints are listed in Table A.2. The weights of the Core controller are listed in Table A.3. The output limits are soft to keep the OCP feasible and to improve numerical stability and solver run time: violations are penalized through the slack variables, with a quadratic weight $q_{s,2}$ and a linear weight $q_{s,1}$, but not excluded. The upper limit on $dp_{\max}$ protects against excessive combustion noise and mechanical loading, which increase with HES.

The embedded implementation must complete each control calculation within the 18 ms window identified in Fig. 3. To meet these tight real-time requirements, the computationally efficient open-source package `acados` is used to solve the embedded NMPC, as depicted in the online controller stage of Fig. 1 [30]. In a simulation study, `acados` achieved shorter computation times than both `MATLAB`'s Model Predictive Control Toolbox using `fmincon` and the `FORCES PRO` [40, 41] backends [23]. This computation-time advantage is attributed to the higher dimension of the state than the control input vector in addition to the short prediction horizon of three engine cycles needed to model the diesel-engine dynamics. This allows the high-performance interior-point method (HPIPM) quadratic-programming solver to take full advantage of the condensation benefits [42, 23, 39, 43]. The plant model can be directly implemented using the discrete dynamics interface of `acados`. The Gauss-Newton approximation is used for the computation of the Hessian in the underlying sequential quadratic programming (SQP) algorithm. The OCP in Eq. 11 produces band-diagonal quadratic-programming subproblems, which are solved using `HPIPM` [42]. The GRU-DNN dynamics model is integrated into the `acados` NMPC implementation on the RPi. This distributed deployment permits more nonlinear-programming and SQP iterations and a deeper dynamics model than the previous long short-term memory-based NMPC on the MABX, while allowing the controller and dynamics model to be updated independently of the main engine-control software [24, 25, 44].

## 3. Experimental Results

All experiments reported in this section were conducted using the embedded real-time implementation on the test bench shown in Fig. 2. The selected NMPC configuration uses balanced cost-function weights for tracking, emission, and mechanical objectives.

The experimental performance of the NMPC over the reference load trajectory of 4,900 engine cycles (approximately 6.5 min) at a constant engine speed of $1500\,\mathrm{min}^{-1}$ is illustrated in Fig. 5. The trajectory is deliberately challenging: it sweeps the full achievable $p_{\mathrm{mi}}$ range (3 to 8 bar), includes rapid step changes and prolonged high-load plateaus that stress the $dp_{\max}$ constraint, and was never seen during DNN training or model-in-the-loop tuning–constituting a genuine validation of both the model and the controller on unseen data. The measured $p_{\mathrm{mi}}$ closely follows the reference; the tracking error spikes during sudden load jumps, partly reflecting the selected actuator-rate versus tracking trade-off. $NO_x$ emissions $c_{\mathrm{NOx}}$ spike during transients and reach up to 971 ppm under high load, while the PM concentration remains below the constraint of $1.5\,\mathrm{mg/m^3}$. As expected, particle-mass concentration $c_{\mathrm{PM}}$ is significantly reduced compared with diesel-only operation, while $c_{\mathrm{NOx}}$ increases. The measured $dp_{\max}$ remains below its specified 15 bar/CAD upper constraint, reaching a maximum of 11.13 bar/CAD during the trajectory. The P2M time $t_{\mathrm{P2M}}$ and the main SOI $\alpha_{\mathrm{Main}}$ vary within narrow bands to shape combustion. Finally, $t_{\mathrm{H2}}$ and HES surge during transients, with HES reaching up to 55.1%, confirming flexible hydrogen/diesel blending. The DES is plotted together with $t_{\mathrm{Main}}$. Overall, the NMPC tracked the reference without measured limit violations while managing the fuel-blend ratio.

Table 5 summarizes these results relative to a diesel-only baseline recorded under identical speed and load conditions without hydrogen injection. The baseline engine was operated by its production controller using the production feedforward calibration map, which serves as a reference condition for the modified, naturally aspirated engine and not as a competing feedback controller. Tracking is reported as MAE, RMSE and NRMSE, the latter normalized by the range of the reference trajectory (5 bar). Using the H2DF controller, the mean absolute error (MAE) for $p_{\mathrm{mi}}$ tracking falls to 0.117 bar (−27.8% vs. diesel) and the RMSE to 0.176 bar (−20.9%), corresponding to a 20.9% improvement in NRMSE. $NO_x$ emissions $c_{\mathrm{NOx}}$, however, increase substantially: the mean value rises by 105.1% to 482.8 ppm relative to diesel, a well-known consequence of H2DF combustion without EGR, as hydrogen's faster flame speed raises peak combustion temperatures [33, 8, 9]. In contrast, $c_{\mathrm{PM}}$ is greatly reduced, with the average dropping by 61.1% to $0.21\,\mathrm{mg/m^3}$ vs. diesel, i.e. roughly 60% given the sensor uncertainty of up to 50% (Section 2.1). $c_{\mathrm{NOx}}$ and $c_{\mathrm{PM}}$ are engine-out concentrations of cylinder 1, not mass-specific emissions: hydrogen displaces intake air, the exhaust flow was not measured, and values in g/kWh cannot be derived from these data. The HES reaches as high as 55.1% and averages 39.7%. If indicated and combustion efficiency are assumed equal to diesel-only operation, the diesel energy displaced by hydrogen equals the HES, so direct combustion-out $CO_2$ is reduced by about 40% on average (mean HES 39.7%) and by up to about 55% at the peak HES. This is a calculated estimate; $CO_2$ and combustion efficiency were not measured.

On the RPi, the NMPC solves in 3 to 7 ms using up to three SQP iterations, well within the 18 ms computation budget available ($t_{\mathrm{calc}}$). No deadline violations or solver failures were observed across all 4900 validation engine cycles.

A key advantage of the NMPC formulation is the direct exposure of its objectives through the cost function weights. By adjusting the relative penalties on $t_{\mathrm{H2}}$ and $t_{\mathrm{Main}}$, the controller can be steered between low and high hydrogen substitution modes without re-identification or re-tuning of the underlying DNN. Table 6 summarizes three configurations: low, mid, and high $H_2$ utilization.

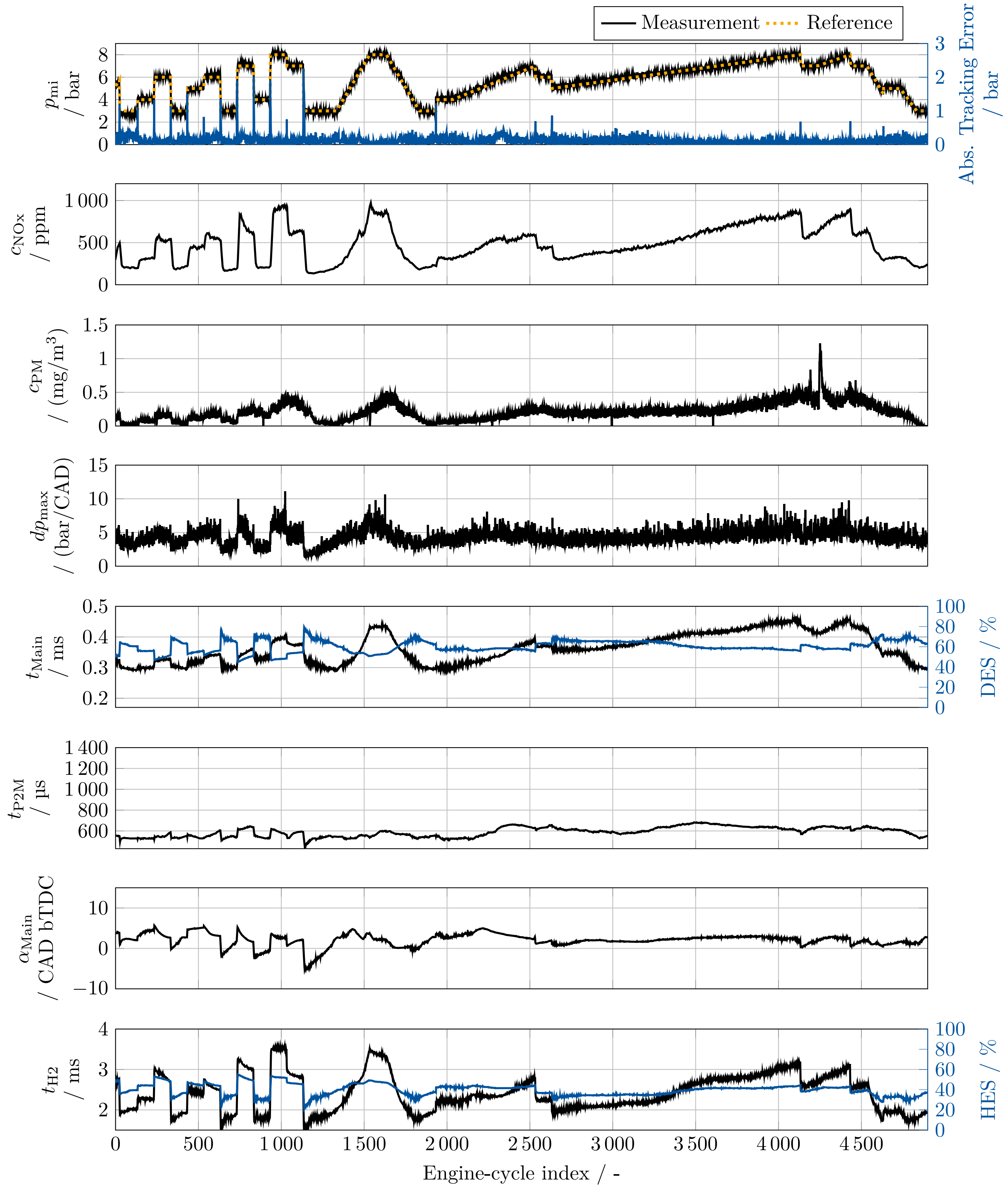

Figure 5: NMPC results for tracking a 4,900-engine-cycle load trajectory at 1500 min$^{-1}$. The plotted quantities are indicated mean effective pressure $p_{mi}$, nitrogen-oxide concentration $c_{NOx}$, particle-mass concentration $c_{PM}$, maximum pressure rise rate $dp_{max}$, main-diesel injection duration $t_{Main}$, pre-to-main interval $t_{P2M}$, main-injection timing $\alpha_{Main}$, hydrogen injection duration $t_{H2}$, diesel energy share (DES), and hydrogen energy share (HES).

The HES reflects the primary difference between the three scenarios, ranging from a mean of 23.1% (*low*) to 44.9% (*high*), with the high-utilization run reaching a peak of 77.8%. As HES increases, $NO_x$ rises monotonically from 417 ppm to 487 ppm mean due to higher peak combustion temperatures. Meanwhile, $c_{PM}$ remains below the soft limit of 1.5 mg/m$^3$ in all runs; given the sensor uncertainty, differences between the weight settings are not interpreted. In all three settings the measured maximum $dp_{max}$ (9.3 to 11.1 bar/CAD) stays below the 15 bar/CAD limit, and no measured violation of any limit occurred.

Table 5: NMPC experimental results for tracking and emissions performance when following the standard load trajectory of 4,900 engine cycles at 1500 min$^{-1}$. Values in parentheses indicate the change relative to the diesel-only production-controller baseline. NOx and PM are engine-out concentrations of cylinder 1; the PM sensor uncertainty is up to 50%.

| Output | Metric | Result (vs. Diesel) |
|---|---|---|
| $p_{mi}$ tracking | MAE / bar | 0.117 (−27.8%) |
| | NRMSE / % | 3.51 (−20.9%) |
| $c_{NOx}$ | max / ppm | 971 (+166.4%) |
| | mean / ppm | 483 (+105.1%) |
| $c_{PM}$ | max / mg/m$^3$ | 1.23 (−45.1%) |
| | mean / mg/m$^3$ | 0.21 (−61.1%) |
| $dp_{max}$ | max / bar/CAD | 11.13 (+101.3%) |
| | mean / bar/CAD | 4.39 (+87.7%) |
| HES | max / % | 55.1 (N/A) |
| | mean / % | 39.7 (N/A) |

Table 6: Comparison of three cost-function weight settings that steer the NMPC towards low, mid, or high hydrogen utilization. All settings are evaluated over the first 4,100 engine cycles of the standard reference trajectory at 1500 min$^{-1}$, which all three runs cover and which span the full 3 to 8 bar load range; the mid-$H_2$ column is the Core run over the same window.

| Metric | | Low $H_2$ | Mid $H_2$ | High $H_2$ |
|---|---|---|---|---|
| **$p_{mi}$ tracking** | | | | |
| MAE | / bar | 0.205 | 0.122 | 0.132 |
| RMSE | / bar | 0.276 | 0.184 | 0.201 |
| NRMSE | / % | 5.53 | 3.68 | 4.02 |
| **$c_{NOx}$** | | | | |
| max | / ppm | 829 | 971 | 1009 |
| mean | / ppm | 417 | 474 | 487 |
| **$c_{PM}$** | | | | |
| max | / mg/m$^3$ | 0.72 | 0.55 | 0.57 |
| mean | / mg/m$^3$ | 0.11 | 0.19 | 0.11 |
| **$dp_{max}$** | | | | |
| max | / bar/CAD | 9.3 | 11.1 | 10.6 |
| mean | / bar/CAD | 4.2 | 4.3 | 4.2 |
| **HES** | | | | |
| mean | / % | 23.1 | 40.2 | 44.9 |
| max | / % | 42.7 | 55.1 | 77.8 |

### *3.1. Robustness Investigation*

Having established the nominal performance and the flexibility of the cost-function weights, this section examines how the controller behaves under perturbations not present during training, quantifying the sensitivity of the embedded GRU-DNN dynamics model to input variations and the closed-loop robustness to measurement noise and model-plant mismatch.

To identify which inputs most strongly influence the embedded model outputs, a local Monte-Carlo sensitivity analysis is performed across a range of baseline conditions and GRU states. Detailed perturbation-level results and pseudo-code are omitted for brevity. Because the inputs are normalized, the dimensionless sensitivity perturbation was set to $\delta_{sens} \in [0, 1]$. The resulting sensitivity matrix is provided in Fig. A.2. It identifies $p_{mi}$ as most sensitive to the hydrogen and main-diesel injection durations, whereas the emissions outputs are less sensitive across the tested perturbation range. This contrast motivates the subsequent closed-loop perturbation of the $p_{mi}$ feedback channel.

The closed loop was then tested on the engine by injecting white Gaussian noise into the $p_{mi}$ feedback channel on the RPi. Fig. 6 shows the response along the standard load trajectory with a noise power of 0.03 bar$^2$ (standard deviation 0.17 bar, peaks of about ±0.5 bar). Even with this unrealistically large injected noise, the NMPC maintained load tracking. Around engine cycle 950, the potential approach to the $dp_{max}$ constraint causes the NMPC to reduce $p_{mi}$ conservatively as a preventive action. At higher-load operating points, where the fluctuating feedback pushed against the $dp_{max}$ constraint, tracking performance degraded. This behavior indicates that the closed loop tolerates strong feedback noise.

To further evaluate the NMPC's robustness, unmodeled dynamics were introduced by varying the plant, specifically $n_{Eng}$. The GRU-DNN dynamics model had been trained using data at a constant engine speed of 1500 min$^{-1}$. Fig. 7 shows an excerpt of a measurement in which the NMPC tracks a smooth reference trajectory varying between 3 bar and 8 bar. In parallel, the engine speed $n_{Eng}$ was varied with smooth ramps between 1200 min$^{-1}$ and 1800 min$^{-1}$ to investigate the NMPC's ability to maintain engine load tracking on $p_{mi}$. The data shows that with increasing engine speed, $NO_x$ emissions slightly decrease due to reduced residence time of air at high temperatures, and conversely increase with reduced engine speed. It is also observable that soot formation increases at higher loads when the engine speed is raised, likely due to reduced time available for mixture formation without corresponding injection advancement. Although the GRU-DNN dynamics model used by the NMPC had not been trained to capture this behavior, the NMPC maintains close $p_{mi}$ reference tracking, with absolute tracking errors remaining below 0.5 bar.

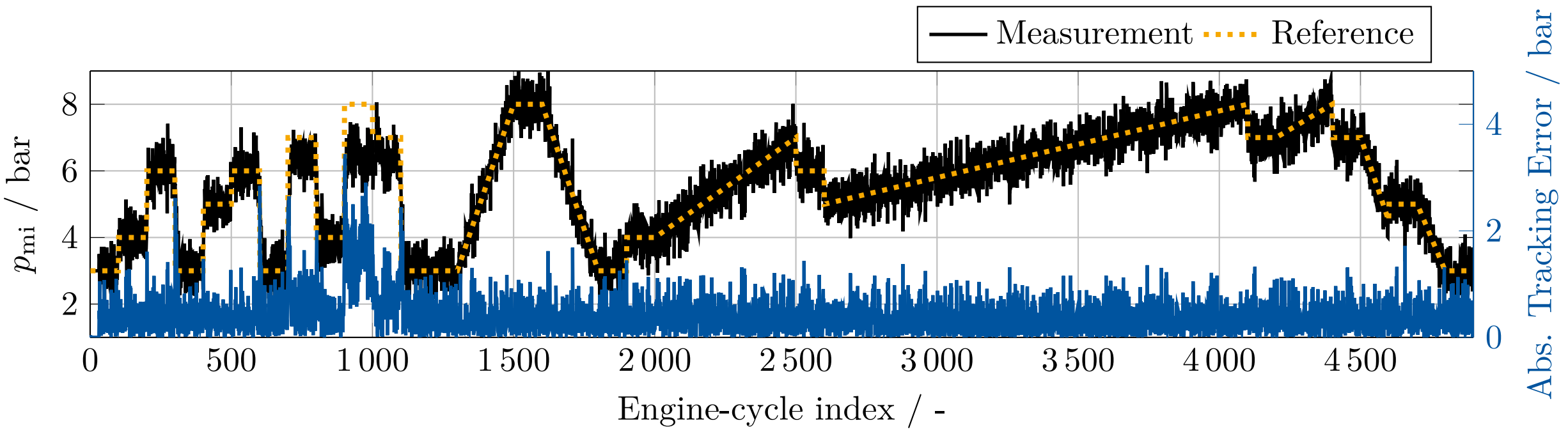

Figure 6: NMPC tracking under high noise injection on the indicated-mean-effective-pressure feedback $p_{\mathrm{mi}}$ over the standard 4,900-engine-cycle load trajectory at $1500\,\mathrm{min}^{-1}$. White Gaussian noise with a noise power of $0.03\,\mathrm{bar}^2$ (standard deviation 0.17 bar, peaks of about $\pm 0.5$ bar) was applied.

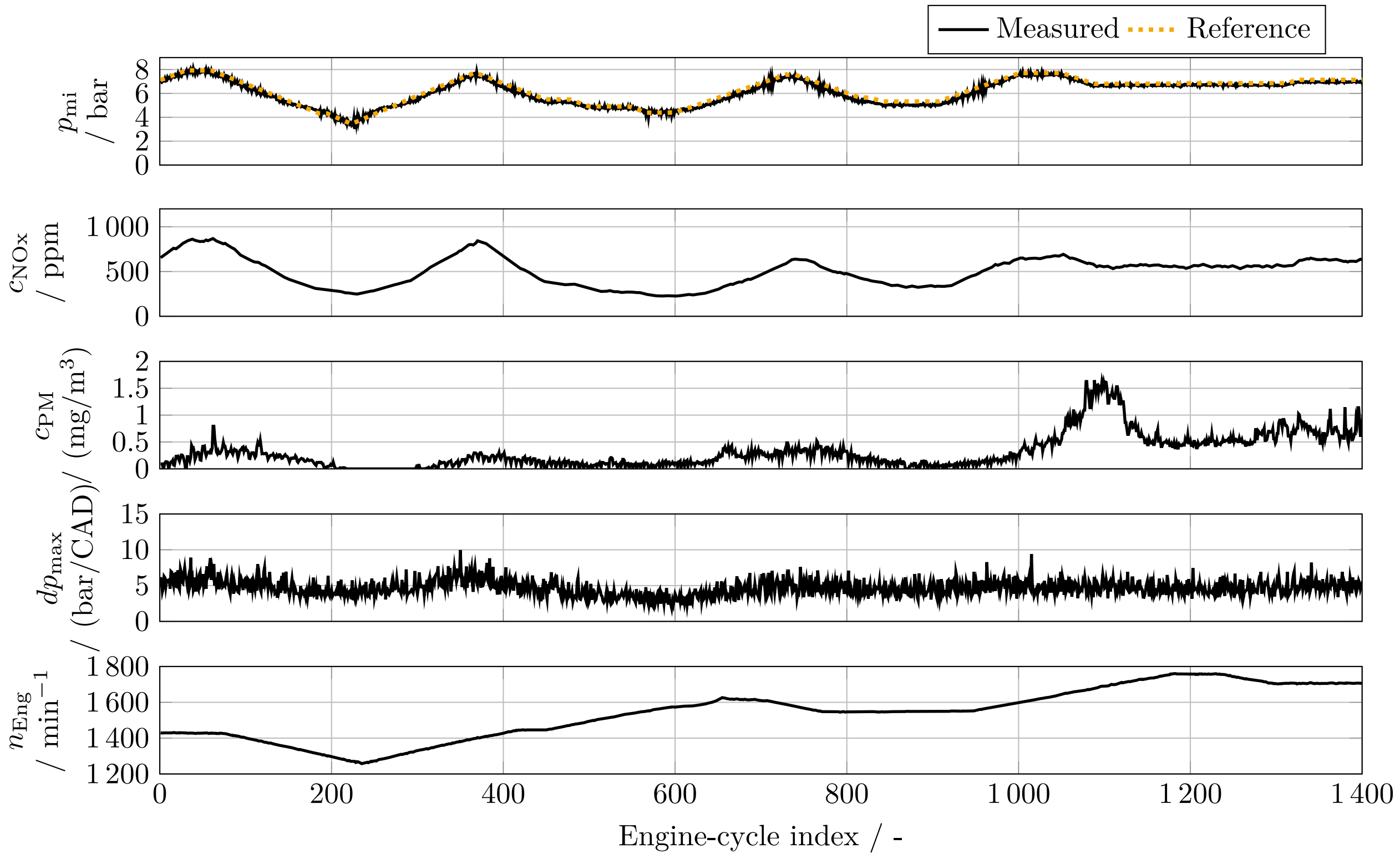

Figure 7: NMPC tracking with model–plant mismatch induced by engine-speed variation $n_{\mathrm{Eng}}$ between $1200\,\mathrm{min}^{-1}$ and $1800\,\mathrm{min}^{-1}$. Outputs are indicated mean effective pressure $p_{\mathrm{mi}}$, nitrogen-oxide concentration $c_{\mathrm{NOx}}$, particle-mass concentration $c_{\mathrm{PM}}$, maximum pressure rise rate $dp_{\mathrm{max}}$, and engine speed $n_{\mathrm{Eng}}$. The actuator trajectories are omitted for brevity.

## 4. Discussion

The main comparison is made against diesel-only operation under the same speed-load conditions. Consequently, the reported mean PM concentration reduction of about 60% reflects the combined effect of hydrogen substitution and NMPC-governed actuator scheduling, rather than an isolated comparison against another H2DF controller at the same mean HES. The increased $NO_x$ emissions are consistent with EGR-free H2DF combustion and are limited by the controller through a soft upper bound rather than eliminated. Likewise, the $CO_2$ benefit discussed here is direct combustion-out $CO_2$ reduction from diesel displacement, estimated from the HES under the assumption of unchanged indicated and combustion efficiency; lifecycle benefit depends on the hydrogen production pathway.

The GRU-DNN dynamics model was trained on pseudo-random-binary-sequence-excited data from the same single-speed operating point, with chronological train-validation-test partitioning to reduce temporal leakage. Although the closed-loop experiments and disturbance tests indicate robust behaviour, the data-driven model has not been validated over a full engine map and does not provide formal stability, robustness or constraint-satisfaction guarantees.

The experiments cover one engine configuration and one run per controller setting. The controller uses $p_{\mathrm{mi}}$ computed from the in-cylinder pressure measurement, which production engines usually do not provide.

## 5. Conclusion

This work presented a GRU-NMPC framework for H2DF engine control, targeting simultaneous load-tracking improvement and multi-objective emission management under safety constraints.

Using 99,800 measured engine cycles, a GRU-DNN dynamics model for $p_{\mathrm{mi}}$, $c_{\mathrm{NOx}}$, $c_{\mathrm{PM}}$, and $dp_{\mathrm{max}}$ was developed first. On held-out test data, the selected model achieved RMSE values of 0.23 bar for $p_{\mathrm{mi}}$, 40 ppm for $c_{\mathrm{NOx}}$, and 1.56 bar/CAD for $dp_{\mathrm{max}}$, with corresponding $\mathrm{NRMSE}_{\mathrm{clip}}$ values of 3.0%, 4.0%, and 12.1%; $c_{\mathrm{PM}}$ prediction achieved 0.14 mg/m$^3$ RMSE and 9.2% $\mathrm{NRMSE}_{\mathrm{clip}}$.

The GRU-DNN dynamics model was integrated into the NMPC using `acados` and deployed on the low-cost RPi embedded target for flexible, modular control execution, as shown in Fig. 1.

The proposed NMPC tracked an unseen, rapidly changing load command over 4,900 engine cycles while partially substituting diesel fuel with hydrogen, achieving a mean absolute error of 0.117 bar (27.8% below the production calibration) and an NRMSE of 3.51% (production calibration: 4.44%). Soot was significantly reduced, with the mean $c_{\mathrm{PM}}$ concentration dropping by about 60% to 0.21 mg/m$^3$, while the mean $c_{\mathrm{NOx}}$ concentration increased by 105.1% to 483 ppm. No measured limit violation occurred; $dp_{\mathrm{max}}$ peaked at 11.1 bar/CAD (limit 15 bar/CAD). Under the balanced weight setting, the HES reached up to 55.1% and averaged 39.7%; a high-$H_2$-utilization weight variant achieved a peak HES of 77.8% and a mean of 44.9%, demonstrating that the framework can be steered towards aggressive hydrogen substitution through cost-function adjustment alone, without model reidentification.

The full controller executed in 3 to 7 ms per engine cycle without deadline violations, demonstrating real-time capability on the embedded target.

Closed-loop engine tests with injected $p_{\mathrm{mi}}$ feedback noise and with engine-speed variation outside the training data, which introduces model–plant mismatch, showed that the NMPC maintains load tracking.

The quantitative results are limited to cylinder 1 of a naturally aspirated four-cylinder engine with hydrogen port injection in this cylinder only, at 1500 min$^{-1}$, without EGR.

Future work includes uncertainty-aware or Lyapunov-based formulations, a feedback-controller baseline and a closed-loop comparison of model structures, state estimation for the GRU state, exhaust-flow measurement for mass-specific emissions, and a systematic study of the weight trade-off. Active follow-on work at the UofA targets a 6.7 L Cummins multi-cylinder H2DF platform and ultimately full-hydrogen operation.

## Funding

This work was supported by the Deutsche Forschungsgemeinschaft (DFG, German Research Foundation) as part of the Research Group FOR 2401 “Optimization-based Multiscale Control for Low Temperature Combustion Engines”. Additional support was provided by the RWTH Aachen University–University of Alberta Junior Research Fellowship, the Natural Sciences and Engineering Research Council of Canada Alliance grant ALLRP 593434-24 and Discovery Grant RGPIN-2024-04990, and Future Energy Systems grant T02-A04. The funding sources had no involvement in the study design; data collection, analysis, or interpretation; manuscript preparation; or the decision to submit the article.

## Acknowledgements

The authors thank the technical staff of the Department of Mechanical Engineering at the University of Alberta for their support during engine test bench operations, especially Javad Kheyrollahi.

## Data and code availability

The software and the measurement data supporting the findings of this study are publicly available on Zenodo:

- H2DF engine-model DNN training software and measurement data: `https://doi.org/10.5281/zenodo.16902580` [45]
- H2DF NMPC software: `https://doi.org/10.5281/zenodo.16902939` [46]

## CRediT authorship contribution statement

**Alexander Winkler**: Conceptualization, Methodology, Software, Investigation, Validation, Writing – original draft, Visualization, Data curation. **Vasu Sharma**: Software, Methodology, Data curation, Writing – review & editing. **Julian Bedei**: Validation, Formal analysis, Writing – review & editing. **Edward Sperling**: Resources, Investigation, Validation, Writing – review & editing. **Charles Robert Koch**: Supervision, Funding acquisition, Writing – review & editing. **David Gordon**: Supervision, Project administration, Funding acquisition, Methodology, Validation, Writing – review & editing. **Jakob Andert**: Supervision, Project administration, Funding acquisition, Writing – review & editing.

All authors have read and approved the final manuscript.

## Declaration of generative AI and AI-assisted technologies in the manuscript preparation process

During the preparation of this work, the authors used Codex 5.6 for grammar, formatting, and language editing. After using this tool, the authors reviewed and edited the content as needed and take full responsibility for the content of the published article.

## Declaration of competing interest

The authors declare that they have no known competing financial interests or personal relationships that could have appeared to influence the work reported in this paper.

## Abbreviations

| Abbreviation | Description |
|---|---|
| CAD | Crank angle degree |
| DES | Diesel energy share |
| DNN | Deep neural network |
| DOI | Duration of injection |
| EGR | Exhaust gas recirculation |
| FC | Fully connected |
| GRU | Gated recurrent unit |
| H2DF | Hydrogen-diesel dual-fuel |
| HES | Hydrogen energy share |
| HPIPM | High-performance interior-point method |
| IMEP | Indicated mean effective pressure |
| MABX | MicroAutoBox |
| MAE | Mean absolute error |
| MPRR | Maximum pressure rise rate |
| NMPC | Nonlinear model predictive control |
| NRMSE | Normalized root mean square error |
| OCP | Optimal control problem |
| P2M | Pre-to-main |
| PM | Particulate matter |
| ReLU | Rectified linear unit |
| RMSE | Root mean square error |
| RPi | Raspberry Pi |
| SOI | Start of injection |
| SQP | Sequential quadratic programming |
| TDC | Top dead centre |
| aTDC | After top dead centre |
| bTDC | Before top dead centre |

## Symbols

| Symbol | Description |
|---|---|
| *Learned model* | |
| $i$, $k$ | Prediction/layer index and engine-cycle index, respectively |
| $u(k)$, $y(k)$ | Physical engine-control input and measured-output vectors |
| $X_{\mathrm{DNN}}(k)$, $Y_{\mathrm{DNN}}(k)$ | Bundled GRU-DNN model input and output vectors |
| $x(k)$, $h(k)$ | Model state and gated-recurrent-unit hidden state |
| $W_\bullet$, $b_\bullet$ | Neural-network weight matrices and bias vectors |
| $z_{\mathrm{FC}i}(k)$ | Output of fully connected layer $i$ |
| $z(k)$, $r(k)$, $\hat{h}(k)$ | Update gate, reset gate, and candidate hidden state |
| $f$, $g$ | Nonlinear state-transition and output-mapping functions |
| $\sigma$, tanh, ReLU | Sigmoid, hyperbolic-tangent, and rectified-linear-unit functions |
| $\odot$ | Element-wise multiplication |
| *Controller formulation* | |
| $\tilde{x}(k)$, $\tilde{y}(k)$ | Augmented NMPC state and output vectors |
| $\tilde{u}_i$ | Manipulated input increment at horizon index $i$ |
| $\Delta u_i$ | Physical engine-control input change at horizon index $i$ |
| $N$ | NMPC prediction-horizon length |
| $\mathcal{I}_N$, $\mathcal{I}_{N-1}$ | State/output and input horizon-index sets |
| $J$, $J_{\tilde{x}}$, $J_{\tilde{u}}$ | Total, state/slack, and input-rate costs |
| $Q$, $R$ | State/output and input-rate weighting matrices |
| $q_i$, $r_i$ | Generic diagonal elements of $Q$ and $R$ |
| $r_{\tilde{u}}$ | Input-increment weighting in the implemented cost |
| $F_{\tilde{u}}$, $F_{\tilde{x}}$, $F_{s,\tilde{x}}$ | Constraint selection, scaling, and slack-mapping matrices |
| $\tilde{u}_{\min}$, $\tilde{u}_{\max}$ | Lower and upper input-increment bounds |
| $\tilde{x}_{\min}$, $\tilde{x}_{\max}$ | Lower and upper augmented-state bounds |
| $s_i$ | Slack vector, with lower and upper state-bound components $s_{\min,\tilde{x},i}$ and $s_{\max,\tilde{x},i}$ |
| *Engine quantities* | |
| $p_{\mathrm{mi}}$, $c_{\mathrm{NOx}}$, $c_{\mathrm{PM}}$, $dp_{\max}$ | Engine outputs: indicated mean effective pressure, nitrogen-oxide concentration, particle-mass concentration, and maximum pressure rise rate |
| $p_{\mathrm{mi,ref},i}$ | Indicated-mean-effective-pressure reference |
| $t_{\mathrm{Main}}$, $t_{\mathrm{P2M}}$, $\alpha_{\mathrm{Main}}$, $t_{\mathrm{H2}}$ | Control inputs: main-diesel duration, pre-to-main interval, main-diesel start angle, and hydrogen duration |
| $n_{\mathrm{Eng}}$ | Engine speed |
| $\alpha_{\mathrm{Pre}}$, $t_{\mathrm{Pre}}$ | Diesel pre-injection timing and duration |
| $p_{\mathrm{Rail}}$, $\alpha_{\mathrm{H2}}$ | Diesel rail pressure and hydrogen start of injection |
| $t_{\mathrm{calc}}$ | Available online calculation time |
| $CO_2$ | Carbon dioxide |
| $NO_x$ | Nitrogen oxides |
| $H_2$ | Hydrogen |

## Supplementary Materials

Table A.1: Training settings for the GRU-DNN dynamics model.

| **Parameter** | **Value** |
|---|---|
| Optimizer | Adam |
| Validation metric | Mean squared error |
| Max. epochs | 5000 |
| Mini-batch size | 512 |
| Verbose frequency | 1 |
| Initial learning rate | 0.001 |
| Learn rate schedule | Piece-wise drop by 25% every 250 epochs |
| L2 regularization | 0.1 |

Table A.2: NMPC constraint bounds used in all experiments, written in the engine-output and control-input symbols of Eq. 12. $\Delta$ denotes the engine-cycle-to-engine-cycle change of the associated input. Soft boundaries are marked with $*$.

| **Lower bound** | **Variable** | **Upper bound** |
|---|---|---|
| 0 bar$*$ | $p_{\mathrm{mi}}$ | 9 bar$*$ |
| 0 ppm$*$ | $c_{\mathrm{NOx}}$ | 1200 ppm$*$ |
| 0 mg/m$^3*$ | $c_{\mathrm{PM}}$ | 1.5 mg/m$^3*$ |
| 0 bar/CAD$*$ | $dp_{\max}$ | 15 bar/CAD$*$ |
| 0.17 ms | $t_{\mathrm{Main}}$ | 0.50 ms |
| 430 µs | $t_{\mathrm{P2M}}$ | 1400 µs |
| −10 CAD bTDC | $\alpha_{\mathrm{Main}}$ | +10 CAD bTDC |
| 1.5 ms | $t_{\mathrm{H2}}$ | 4.0 ms |
| −0.005 ms/engine cycle | $\Delta\, t_{\mathrm{Main}}$ | 0.005 ms/engine cycle |
| −20 µs/engine cycle | $\Delta\, t_{\mathrm{P2M}}$ | 20 µs/engine cycle |
| −1.5 CAD/engine cycle | $\Delta\, \alpha_{\mathrm{Main}}$ | 1.5 CAD/engine cycle |
| −0.24 ms/engine cycle | $\Delta\, t_{\mathrm{H2}}$ | 0.24 ms/engine cycle |

Table A.3: NMPC cost weights of the Core controller in Eq. 12. All quantities are normalised, so the weights are dimensionless. Eq. 12 shows the terms for $t_{\mathrm{Main}}$ and $t_{\mathrm{H2}}$; the same form applies to $t_{\mathrm{P2M}}$ and $\alpha_{\mathrm{Main}}$. The weights on the GRU hidden states are zero. The values are those set in `acados`, whose cost carries a factor $\frac{1}{2}$ on all quadratic terms, including $q_{s,2}$, but not on the linear slack term $q_{s,1}$.

| **Cost term** | **Weight** | **Value** |
|---|---|---|
| Load tracking, $p_{\mathrm{mi}}$ | $q_{p,\mathrm{mi}}$ | 1 |
| Emissions, $c_{\mathrm{NOx}}$ | $q_{c,\mathrm{NOx}}$ | $10^{-3}$ |
| Emissions, $c_{\mathrm{PM}}$ | $q_{c,\mathrm{PM}}$ | $10^{-3}$ |
| Pressure-rise rate, $dp_{\max}$ | $q_{dp,\max}$ | 0.1 |
| Main injection duration, $t_{\mathrm{Main}}$ | $q_{t,\mathrm{Main}}$ | $6 \times 10^{-4}\,*$ |
| Pre-to-main spacing, $t_{\mathrm{P2M}}$ | $q_{t,\mathrm{P2M}}$ | $10^{-3}$ |
| Main injection timing, $\alpha_{\mathrm{Main}}$ | $q_{\alpha,\mathrm{Main}}$ | $10^{-3}$ |
| Hydrogen duration, $t_{\mathrm{H2}}$ | $q_{t,\mathrm{H2}}$ | $10^{-3}\,*$ |
| Input rate, all four inputs | $r_{\tilde{u}}$ | 0.1 |
| Slack, linear and quadratic | $q_{s,1}$, $q_{s,2}$ | 10 |

$*$ Changed in the low- and high-$H_2$ runs of Table 6: the weights on $t_{\mathrm{Main}}$ and $t_{\mathrm{H2}}$ were varied to steer the controller towards lower or higher hydrogen utilisation.

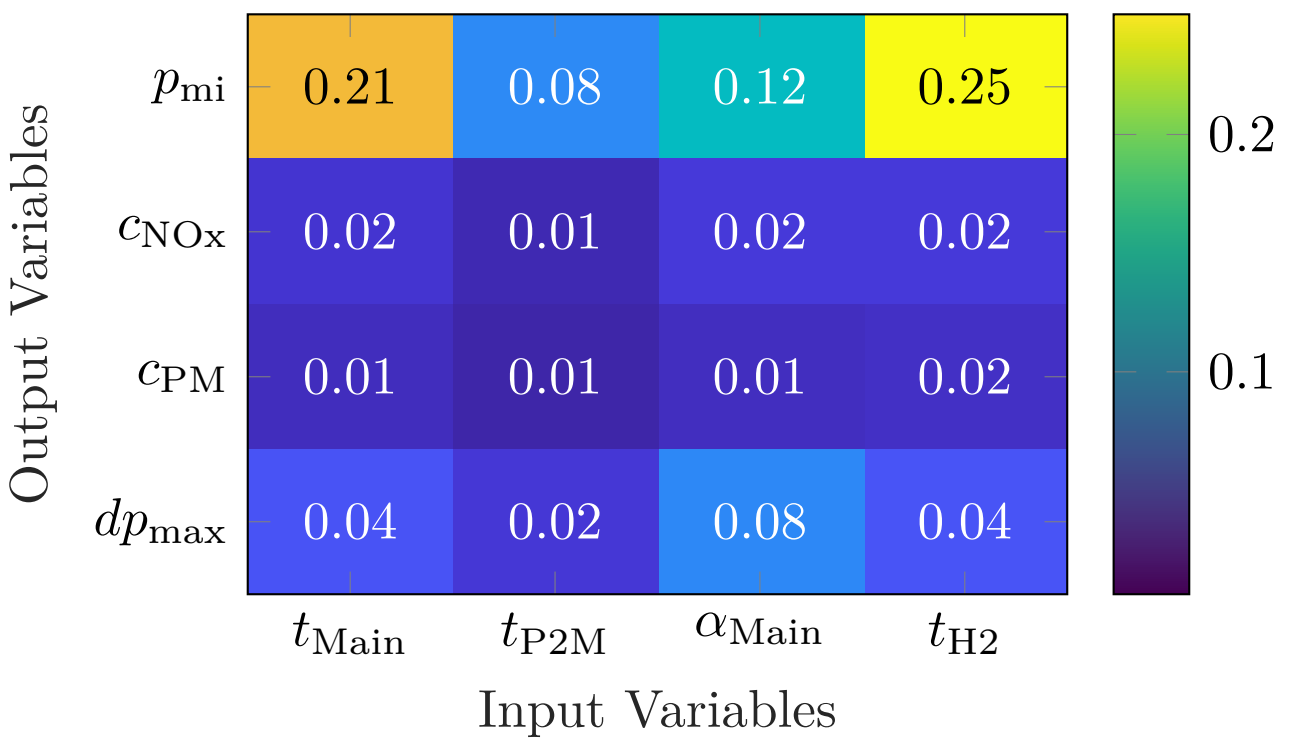

Figure A.2: Sensitivity matrix of the GRU-DNN dynamics model. Inputs are main-diesel duration $t_{\mathrm{Main}}$, pre-to-main interval $t_{\mathrm{P2M}}$, main-diesel start angle $\alpha_{\mathrm{Main}}$, and hydrogen duration $t_{\mathrm{H2}}$; outputs are indicated mean effective pressure $p_{\mathrm{mi}}$, nitrogen-oxide concentration $c_{\mathrm{NOx}}$, particle-mass concentration $c_{\mathrm{PM}}$, and maximum pressure rise rate $dp_{\max}$.

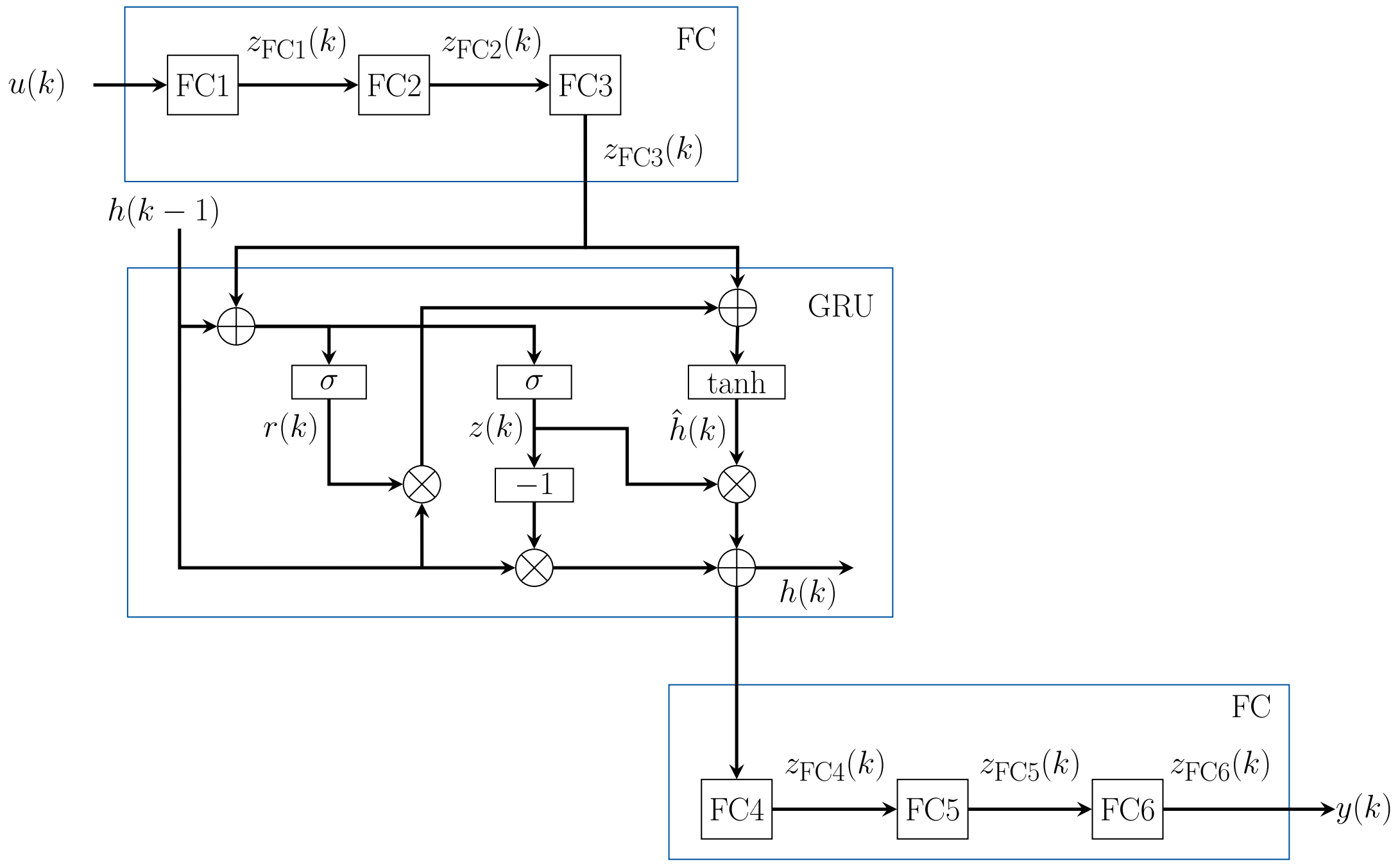

Figure A.1: Computational graph of the deployed GRU-DNN dynamics model. Fully connected (FC) layers use rectified linear unit (ReLU) activations; $u(k)$ and $y(k)$ are the control and output vectors, and the GRU cell propagates the hidden state $h(k)$ between engine cycles.

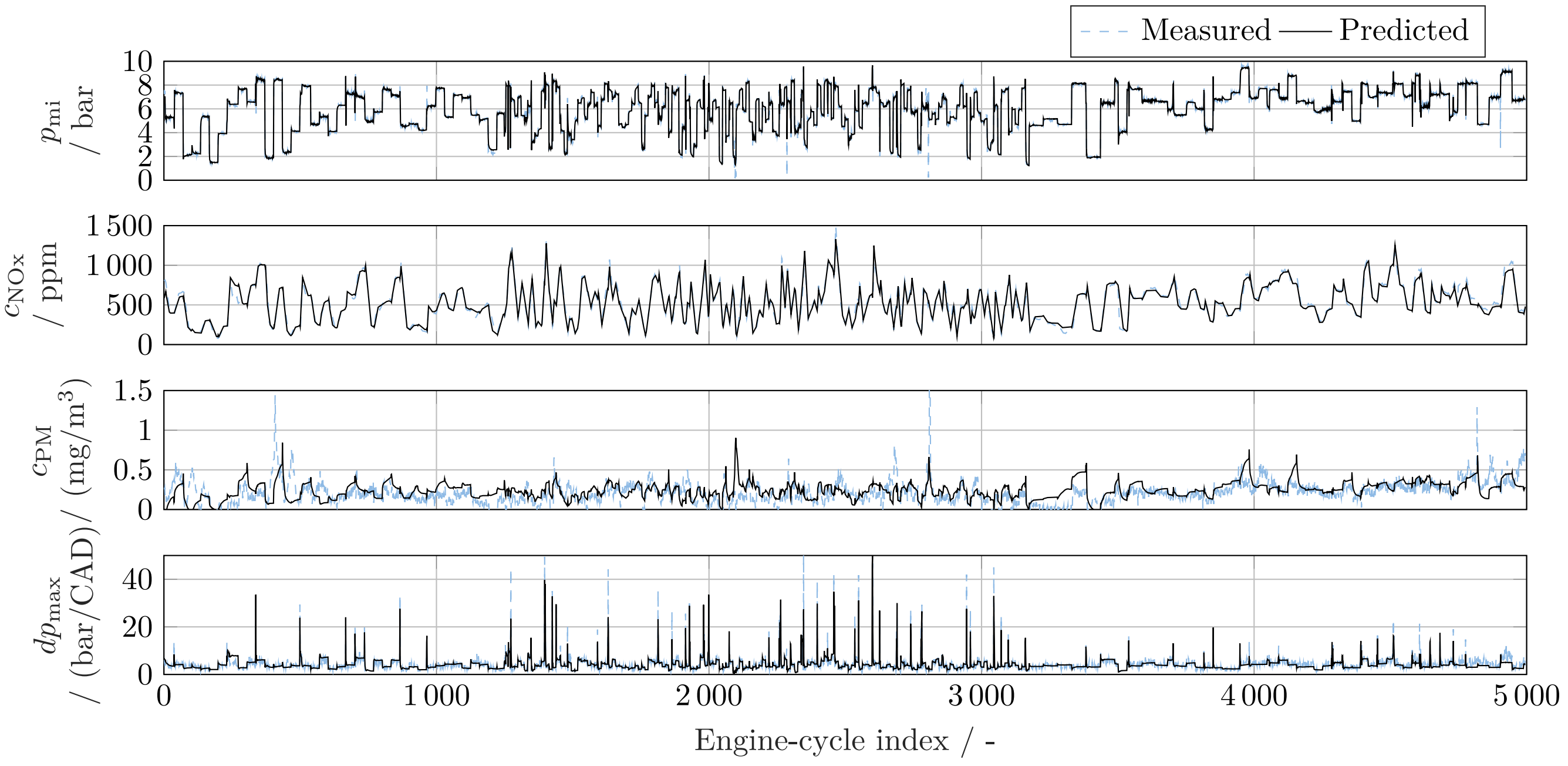

Figure A.3: Measured and predicted GRU-DNN outputs on the held-out test data: indicated mean effective pressure $p_{\mathrm{mi}}$, nitrogen-oxide concentration $c_{\mathrm{NOx}}$, particle-mass concentration $c_{\mathrm{PM}}$, and maximum pressure rise rate $dp_{\mathrm{max}}$.